\documentclass[aps,prb, preprint]{revtex4-1}
\usepackage{graphicx}
\begin{document}
\title{Evidence for incompressible states in a metal graphene tunnel junction in high magnetic field}
\author{C. E. Malec and D. Davidovi\'c}

\affiliation{School of Physics, Georgia Institute of Technology,
Atlanta, GA 30332}

\date{\today}

\begin{abstract}
We present transport measurements of tunnel junctions made between Cu and graphene in a magnetic field.  We observe a transition to a Landau level like structure at high fields, as well as a set of sharp features in the tunneling spectra that shift with gate and tunnel probe voltage along the lines of constant charge density. We explain the sharp features with the formation of degeneracy split localized Landau levels, and addition of electrons to those levels one by one. A large capacitive coupling to the tunnel probe also increases the gate voltage spacing between the Landau levels.
\end{abstract}
\maketitle

The electron and hole bands of pristine graphene exhibit linear dispersion, and meet at a point in reciprocal space of zero energy and vanishing density of states (DOS) known as the Dirac point.  This situation leads to charge carriers that behave as massless relativistic particles, and a large number of theoretical and experimental studies have investigated the consequences of this fact.
For review see Refs.~\cite{geim2,castroneto1}.

Of increasing importance for future graphene technology is the interaction between graphene and a metal.  Since metals will eventually be involved in some way in graphene circuits, as interconnects or gates, it is critical to understand their effect on the graphene.  Theoretical~\cite{giovannetti,barraza} and experimental~\cite{huard,venugopal,pi,xia,malec,amet} publications have found that one of the major effects of a metal on graphene is doping.  In addition to this, it is believed that some metals interact weakly, leaving the linear dispersion of graphene undisturbed, and thus it may be desirable to use such metals near regions where graphene properties are most important.

In this work, Cu/graphene tunnel junctions are used to investigate the structure of the field dependent graphene density of states (DOS) beneath a metal contact near the Dirac point.  We can also identify certain aspects of the data which can only be explained by the more complex physics of single electron charging, providing physical evidence for the strongly localized Landau levels which play a crucial role in the quantum Hall effect.

Previous studies of tunneling in graphene in magnetic field have been performed with Scanning Tunneling   Spectroscopy (STS)~\cite{li, jung, luican}, however these studies cannot study graphene in proximity to a bulk metal.  STS has also been performed on graphene on metal substrates such as Cu~\cite{gao2} and Au~\cite{tao}, but the lack of a gate electrode or magnetic field in these studies creates difficulty in interpreting peaks in the dI/dV spectrum.  Finally, solid state devices in a geometry similar to our own have been performed with a superconducting Pb electrode.~\cite{dirks}  This study also claims, as we will later, that a localized region is created underneath the metal electrode.  However, the study investigates excitations within the superconducting gap region in zero field, whereas our present study concerns tunneling between a normal metal and graphene in the presence of up to 12T.

In order to fabricate our tunneling junctions, graphene is first mechanically exfoliated from single crystal graphite obtained from Naturally Graphite\texttrademark~ onto a highly doped Si wafer with a 300nm layer of thermally grown oxide,~\cite{novoselov0} and then single layers are confirmed by Raman spectroscopy~\cite{ferrari}.  Cu junctions were directly evaporated onto graphene by e-beam lithography.  After lift-off in acetone, the device is placed in dilute nitric acid ($\sim$12\% HNO$_3$) for several seconds, and then rinsed with DI water, isopropanol, and blown dry.  The acid promotes oxidation of the Cu, and some junctions become highly resistive after this treatment ($>$1 M$\Omega$), afterwards they continue to age, though slowly, taking several weeks to fully oxidize.  At low temperatures, the resistance values are stable, as the oxidation rate is greatly suppressed. See Ref.~\cite{malec} for additional details about fabrication.

Narrow junctions ($<$ 200 nm) were used to insure that the acid could partially penetrate the Cu/graphene interface before the lead was overetched.  Though slight tarnishing could be seen in the leads, only leads that were highly conductive at low temperatures (k$\Omega$s) were used as a grounding electrode.  The conductance of the tunnel probe was not severely suppressed at low temperatures, meaning that the junction between metal and graphene provided the primary mode of resistance in the circuit, and not a discontinuity in the metal lead.  Also, as we shall show, the tunneling spectra exhibit expected graphene properties.

\begin{figure}
\begin{center}
\includegraphics[width=.99\textwidth]{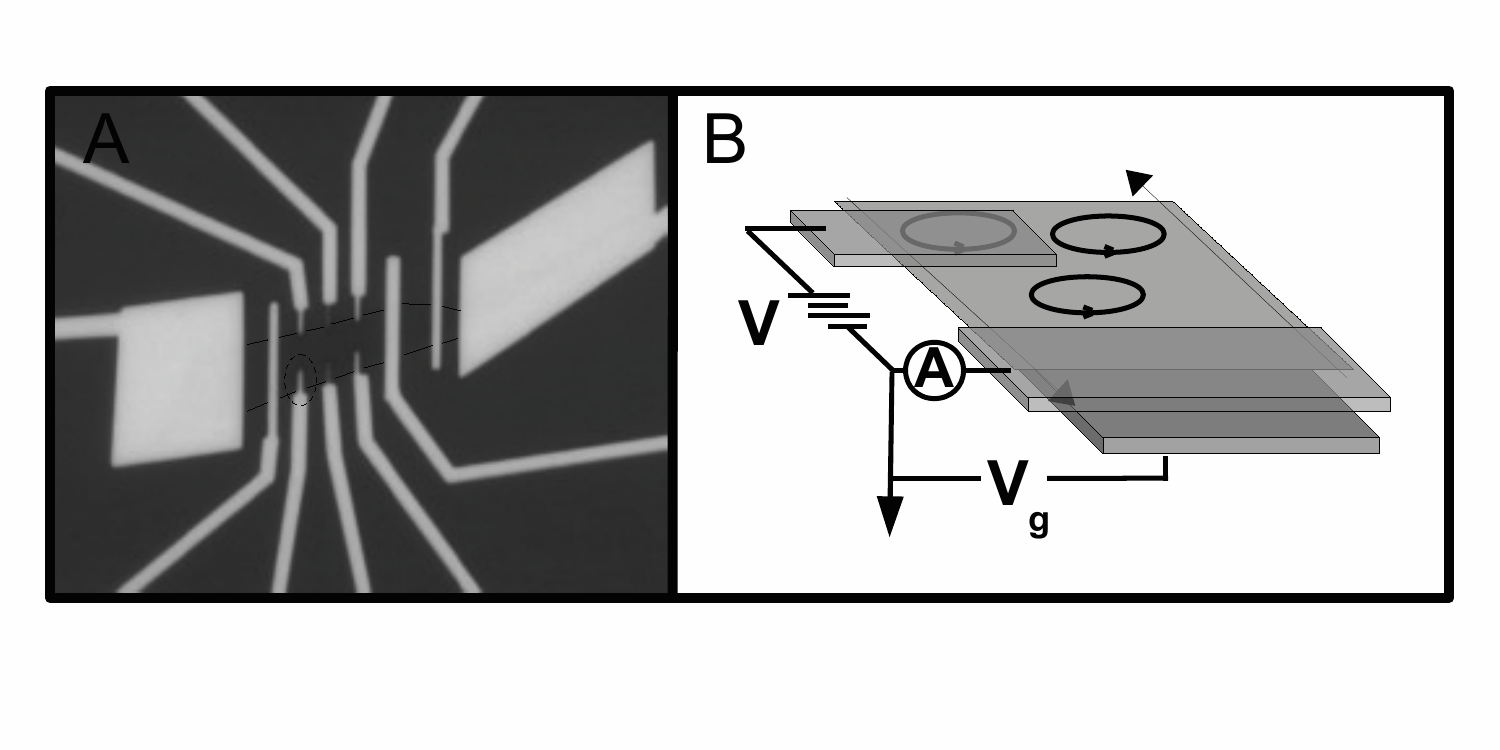}
\caption{A: An optical picture of the graphene used in this experiment, the tunnel probe is circled with a dotted line.  B: A cartoon of the experimental setup showing the tunnel probe with a localized and extended state beneath it.}
\label{device}
\end{center}
\end{figure}

We report on one junction in particular, however the primary results presented here were observed in two other devices that displayed tunneling properties.  Fig.~\ref{device}-A shows an optical image of the completed device, with a dotted line encircling the particular probe used in data discussed.  Many of the other probes on this device did not become highly resistive after the short exposure to HNO$_3$.  In the present case the junction was only partially overlapping the sample, and in two other cases the junction was entirely overlapping.  All measurements presented here were taken while the sample was in vacuum in a cryostat operated at 2.2 K, and equipped with a superconducting magnet.  Tunnel spectra were gathered by applying a DC probe voltage, $V_p$, across a highly resistive and an ohmic probe, and a gate voltage, $V_g$, between the Si back gate and an ohmic probe, and measuring the current as shown in Fig.~\ref{device}-B.  A numerical derivative was then taken to calculate the conductance.

\begin{figure}
\begin{center}
\includegraphics[width=.75\textwidth]{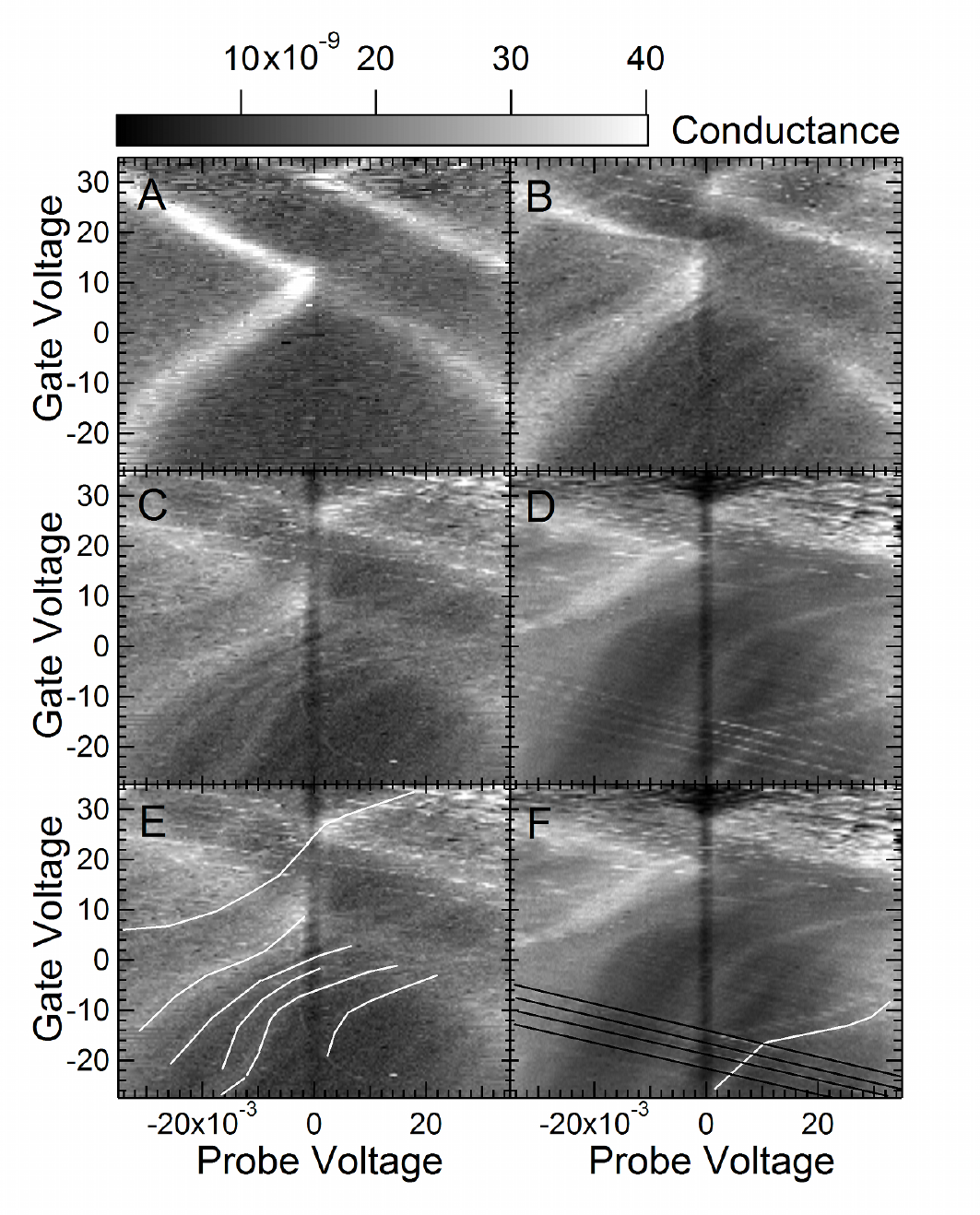}
\caption{A-D: Progression of conductance maps from 0 to 12T.  The original set of resonances, due to the disorder potential, are broken up and replaced by a series of staircases.  E-F: The same conductance maps as C \& D, with white guidelines to mark the formation of staircase structure in the resonances, and black lines to show the formation of structure associated with the charging of localized states.  The white line in F decreases in slope exactly at the point where it crosses the black line showing that the DOS drops after filling the Landau level, consistent with ~\cite{jung}}
\label{data}
\end{center}
\end{figure}

Our data consists of mapping the conductance versus  $V_g$ and $V_p$ at different magnetic fields.  Fig.~\ref{data} displays our conductance maps at 0, 6, 9, and 12T.  The tunnel spectra are dominated by the gate voltage independent zero bias anomaly (ZBA), similar  to other tunneling experiments.~\cite{jung,zhang3}  We can analyze the data despite the ZBA by inspecting how tunnel resonances in the spectra shift with $V_g$.  Those features that shift with gate voltage are related to the graphene DOS at energy $eV_p + E_F(V_p,V_g)$, and their slope is directly proportional to the DOS,~\cite{malec} where $E_F(V_p,V_g)$ is the Fermi level in graphene.  By applying a gate or probe voltage, charge is added to or subtracted from the graphene.  In regions of high DOS, the added charge does not shift the Fermi level much, whereas in regions of low DOS, the added charge shifts the Fermi level quickly.

Looking at Fig.~\ref{data}-A, lines of high conductance can be seen at 0T.  These are resonances in the DOS caused by the disorder potential, and the slope $dV_g/dV_p$ becomes slightly flatter at positive gate voltage, where the density of states is lower.  This we interpret as a broadened Dirac point,~\cite{malec} and its location at positive gate voltage means the graphene is hole doped.  As the magnetic field is increased, these lines are seen to break up, and are replaced by lines with a staircase like structure of alternating high and low sloped lines. The Staircase like structure is indicated by the guides in Fig.~\ref{data}-E and F.  We identify these lines as the formation of Landau levels.  Since our junction covers a large area compared to the expected magnetic length in graphene, as well as the charge puddles caused by inhomogeneous doping, it is possible that the conductance maps represent conductance from multiple channels, and so features in addition to those predicted by theory may be present.

To compare our conductance maps with those expected from perfect graphene we numerically calculate the tunneling conductance assuming constant transmission with respect to energy for states between the Fermi level and the probe voltage for given magnetic field, probe, and gate voltages.  The semiclassical model for the ideal Landau level structure of graphene gives the DOS in a magnetic field as $\sum_{i=-\infty}^{\infty}\frac{4eB}{h}\delta\left(E-v_Fsign(i)\sqrt{2e\hbar B|i|}\right)$.  The use of such a DOS is justified by analytic~\cite{schnez} and numeric~\cite{libisch} calculations of the electronic spectrum for confined graphene quantum dots in a magnetic field, which show that the localized states of the quantum dot still converge upon the expected landau level spectrum at high magnetic fields.

\begin{figure}
\begin{center}
\includegraphics[width=.75\textwidth]{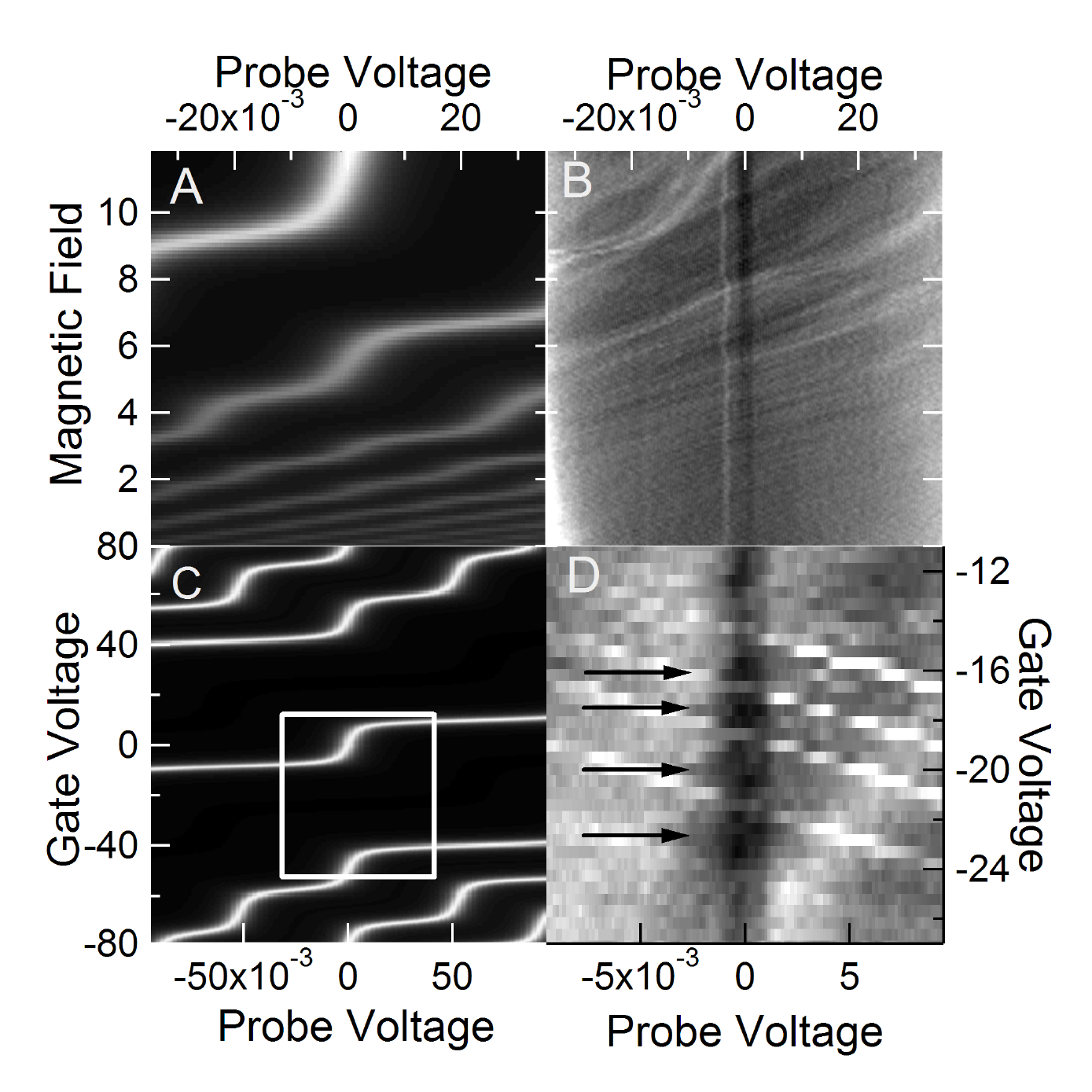}
\caption{A: A simulation of a $V_p$ vs B map assuming undoped graphene and -35 V on the back gate.  B: Data taken at 0V gate voltage in the sample under discussion.  C: A simulation of a $V_p$ vs $V_g$ map, showing the staircase structure of the Landau levels at 12T, a box denotes the approximate range of at least one of the conductance channels measured in our data.  D: Arrows mark the Coulomb diamonds at 12T.}
\label{simulations}
\end{center}
\end{figure}

To convert from the DOS to the conductance map, we use the capacitor model described in Ref.~\cite{malec} to solve for $E_F(V_g,V_p)$. We arrive at Fig.~\ref{simulations}-C by applying a 3.5meV Lorentzian broadening function to the ideal DOS and using a capacitance ratio between the tunnel probe and the back gate's capacitance per unit area of 250.  A staircase structure can be seen to emerge in which the gate voltage is unable to shift features in the spectra until the highly degenerate Landau levels are filled, and then the spectra are quickly shifted when the DOS is low.

We can use the same assumptions to simulate conductance maps vs. $B$ and $V_p$.  Ideally, such a map should show a series of parabolas shifted from the center depending on the doping level and applied gate voltage.  The simulation of a conductance map vs. $B$ and $V_p$ in Fig.~\ref{simulations}-A shows that the expected parabolic lines are disrupted by shifts in the Fermi level caused by the tunnel probe. We show for comparison data taken from our tunnel probe which shows similar behavior (Fig. 3B), at $V_g=0V$.  
  
An important observation of this paper is the sharp resonances that begin to appear in the upper portion of the 6T map as seen in Fig.~\ref{data}-B, and become very well resolved, with the lowest resonance appearing at $V_g=-20V$ and $V_p=0 mV$ at 12T in Fig.~\ref{data}-D.  In prior work,~\cite{malec} we showed that the low carrier density regions in graphene, produce anomalies in the tunnel conductance maps that shift along the lines of constant charge density $C_pV_p+C_gV_g=const$.  The resonance shifts in Fig. 2B-D can be attributed to low density regions with $C_p/C_g=250$.

We interpret these resonances to be due to single electron charging effects on localized states near the filling of Landau levels.  In the progression from Fig.~\ref{data}-B-D, the set of sharp resonances where the lowest resonance appears near $+25$V do not appear to shift, whereas the second set of resonances is seen to shift from near 0V in Fig.~\ref{data}-C to $\sim$-20V in Fig.~\ref{data}-D.  This, combined with its clear appearance before any other level, leads us to the conclusion that the Dirac point is near +25V, and that these resonances are associated with the n=0 Landau level.  The other set of resonances which show near -20V at 12T would therefore be the n= -1 Landau level.  The bright, high conductance, positive sloping lines at -20V, indicated by white guides to the eye in Fig.~\ref{data}-F, changes slope after intersecting with the top most resonance line. The change in slope is consistent with that found in Ref.~\cite{jung}, which is another indicator that the resonances form when the LL level is filling.

The gate voltage spacing is much larger than expected if these are in fact the n=0 and n=-1 levels.  The spacing between the bottom resonance of the n=-1, and the bottom resonance of the n=0 is $\sim$45 V, when 16 V is expected at 12T.  A possible explanation is a capacitive division due to the tunnel probe, and numerical solution of the electrostatic equations shown in Fig. 3-C demonstrate that the spacing in gate voltage between the various Landau Levels is nonuniform.  Since graphene has a low density of states compared to that in a metal, the electric potential of the graphene can shift without a corresponding change in the electrochemical potential of the grounding electrode as long as the Fermi level of the graphene compensates such that $\mu_F+e\phi=0$. In ideal graphene in a magnetic field, the density of states between the Landau levels is zero, and the electric potential follows capacitive division between the gate and the probe potential.

The required gate voltage to cause a given shift in graphene's electric potential can be solved using the equation $C_g(\Delta V_g-\Delta\phi)+C_p(-\Delta\phi)=-\Delta Q$ where $\Delta Q$ is the induced charge on graphene, $\Delta\phi$ is the electric potential shift in graphene under the probe, and the tunnel probe potential is held at zero.  To shift from the n=-1 to n=0 Landau level requires first that the electric potential is shifted such that the Landau level to be filled aligns with the electrochemical potential of the leads meaning that $-e\Delta\phi=\Delta\mu_F=v_F\sqrt{2\hbar B}=\Delta\epsilon_{LL}$, and second that the necessary charge density to fill a Landau level, $-4e^2B/h$, is supplied from the leads.  Therefore, $e\Delta V_g=\frac{(C_p+C_g)}{C_g}\Delta\epsilon_{LL}+\frac{4e^2B}{h}$.  The first term on the right is small when $C_p$ is small ($\sim 125mV$ at 12T) and can typically be ignored, but if $C_p/C_g=250$ then this term becomes 31$V$, and must be taken into account.  Thus $\Delta V_g=47V$ as in our data, rather than the expected 16$V$.  The resonances near 0V in Fig. 2-C are not well developed enough to accurately measure their position.


A closer inspection of Fig.~\ref{simulations}-D reveals the resonances are spaced by approximately 10meV along the $V_p$ direction.  The fact that the spacing is much larger than the expected Zeeman (1.4meV) splitting in the 1st Landau level of graphene, indicates that the charging energy of the Landau level dominates these spacings. It vaguely appears that the tunnel conductance of the sample, near zero probe voltage, is suppressed within the region enclosed by four diamonds, indicated by the arrows in Fig.~\ref{simulations}-D, suggesting that the resonances can be explained by the theory of Coulomb blockade.~\cite{averin}  In the present case, a localized state within a Landau level becomes the central island, tunnel coupled to both the metal lead and an extended state within the graphene which in turn couples to the grounding electrode.~\cite{jung} Such samples exhibit sudden onsets of current along the sides of a Coulomb diamond.  The two negative slopes of the Coulomb diamond ($\Delta V_p/\Delta V_g=-C_p/C_g$) occur when the island is at a large enough potential for an electron to tunnel into (out of) the extended state, followed by an electron tunneling off (onto) the probe. The spacing between the negative sloping lines, measured parallel to the probe voltage axis, is $e/AC_p$, where $A$ is the area of the island, assuming the energy level spacing due to Zeeman and the valley degeneracy splitting much smaller than the charging energy.

We can derive the same expression for the spacing more generally, without using the theory of Coulomb blockade, which is important because the Coulomb diamonds are rather vague. As discussed earlier, the regions of low density in the conductance maps versus $C_p$ and $C_g$, shift along the lines of constant charge density $q=C_pV_p+C_gV_g$. According to our interpretation, moving from one line to the next one, along the direction of increasing $V_p$ ($V_g$),  corresponds to adding an electron to an unoccupied, degeneracy split LL. Thus, $q$ to changes by $e/A$ from line to line, leading to the spacing $e/AC_p$ ($e/AC_g$).
 
 Thus, we obtain $AC_p=16aF$.  Assuming a gate capacitance per unit area of $124\mu F/m^2$ as measured on a test wafer from the same batch of oxidized Si wafers, as well as $C_p/C_g$ as described earlier, we calculate a radius of the localized state of 13 nm. The magnetic length of graphene at 12T is 7.4nm~\cite{neto}, which suggests that tunneling could indeed be occurring through a single localized Landau level.  We note that the resonances located near +25V gate voltage have a larger spacing ($\sim$15meV).  As the region of graphene affected by the gate becomes much smaller than the 300nm gate dielectric thickness, the breakdown of the plane capacitor model can explain small discrepancies in the measured and expected length scales.

Our data is to be compared with, ~\cite{jung} in that work, the Landau levels are identified as a staircase structure in the conductance maps, and at high fields, sharp resonances are seen to emerge at the filling and emptying of Landau Levels.  It was suggested that the larger energy spacing near the zero Landau level was due to a lifted valley degeneracy, adding 10meV at 8T to the charging energy in the n=0 Landau level.  The staircase structure has recently also been observed in ref.~\cite{luican}.

In conclusion, transport through graphene tunnel junctions in a high magnetic field demonstrates localized states in the quantum Hall regime. The direct probing of a localized state in a Landau level is both important as confirmation of earlier experiments, and a new way to study Landau level physics in simple solid state device. Also, the capacitive coupling between the Landau level and the probe electrode increases the apparent spacing between the Landau levels in gate voltage, and should be accounted for when assigning energy spacings to Landau levels in spectroscopic studies.
\\\\
{\bf Acknowledgements}\\
We would like to thank Markus Kindermann, Salvador Barraza-Lopez, and the reviewers for helpful comments.  Funding provided by DOE grant DE-FG02-06ER46281

\newpage

\end{document}